# Aperiodic topological order in the domain configurations of functional materials

Fei-Ting Huang and Sang-Wook Cheong

Rutgers Center for Emergent Materials and Department of Physics and Astronomy, Rutgers University, Piscataway, New Jersey 08854, USA

Correspondence to SWC: sangc@physics.rutgers.edu

**In numerous functional materials, such as steels, ferroelectrics and magnets, new functionalities can be achieved through the engineering of the domain structures, which are associated with the ordering of certain parameters within the material. The recent progress in technologies that enable imaging at atomic-scale spatial resolution has transformed our understanding of domain topology, revealing that, along with simple stripe-like or irregularly shaped domains, intriguing vortex-type topological domain configurations also exist. In this Review, we present a new classification scheme of '$Z_m \times Z_n$ domains with $Z_l$ vortices' for 2D macroscopic domain structures with $m$ directional variants and $n$ translational antiphases. This classification, together with the concept of topological protection and topological charge conservation, can be applied to a wide range of materials, such as multiferroics, improper ferroelectrics, layered transition-metal dichalcogenides and magnetic superconductors, as we discuss using selected examples. The resulting topological considerations provide a new basis for the understanding of the formation, kinetics, manipulation and property optimization of domains and domain boundaries in functional materials.**

Furthering our understanding of and developing control over domains and domain boundaries are essential for the identification of the origin of macroscopic physical properties of functional materials and to enable their exploitation in technological applications[1-10]. The ordering of the charge, spin, orbit and chiral degrees of freedom, which are often relevant for materials functionality, accompanies the formation of domains and domain boundaries associated with directional variants[11-17] and translational antiphases[18-20]. Translational antiphases overlap with each other through a translation symmetry operation; all other symmetry operations, including rotational operations, are relevant to directional variants. Many of these domain structures form in either irregular or simple geometric shapes, such as stripes[3,18,21-24]. However, over the past few years impressive developments in imaging techniques that enabled the investigation of physical properties with high spatial resolution have changed our understanding of domain structures (Table 1). For example, a recent Lorentz electron microscopy experiment employing in-situ femtosecond laser excitations remarkably revealed optically-induced magnetic vortex-antivortex networks in iron thin films, which was compared with the universal so-called Kibble-Zurek mechanism for the generation of topological defects[22]. Ferroelectric domains, which often show rectilinear domains, were considered by the mid-twentieth century to be well understood. However, intriguing cloverleaf-like ferroelectric domains that exhibit improper ferroelectricity and novel magnetic properties[25,26] and that are similar to the well-known schlieren textures in liquid crystals[27-28] (http://www.nsf.gov/news/mmg/mmg_disp.jsp?med_id=59511) were unveiled in hexagonal rare-earth manganite ($h$-$RMnO_3$) in transmission electron microscopy (TEM)[14,29-31] and various scanning force microscopies[14,32-35] experiments. The discovery of these beautiful ferroelectric cloverleaf domains with vortex structures provided a new paradigm in the quest for novel multifunctional materials, because these domains exhibit lattice distortions, ferroelectricity, magnetism and electronic conduction[14,26,36-39].

$h$-$RMnO_3$ was initially considered to be a unique example of a material hosting such cloverleaf domains, as solid materials are, in general, not expected to preserve this kind of vortex-like objects, which are analogous to the vortices realized in soft matter systems such as liquid crystals[27-28] and superfluids[40,41]. As it turned out, this is not the case, and various vortex-like domain structures with aperiodic topological order have been observed over the past few years in a number of bulk solid materials and in thin films (Table 1). These functional materials can exhibit different orientations of the directional orders, thus spatially varying vectors (order parameters) are accommodated through delicate vortex structures. These emergent vortex-like domain configurations in real space are essential for the understanding of the energy hierarchy of different types of domain boundaries[38,42,43]. Another milestone was the discovery of emergent physical properties — entirely distinct from those of the domains — at the domain boundaries. These properties, which include local conduction[2,3,32,33,44,45], enhanced electromechanical responses[46] and a ferromagnetic net moment[34,47] (whereas the domains exhibit standard antiferromagnetic order or insulating behaviour) have sparked a broad interest in nanoscale domain boundary engineering for memory and spintronic applications[1]. A useful classification of domains and domain vortices can be developed based on their topology, using cyclic groups — algebraic groups generated by a single element from which every other element in the group can be obtained by applying an appropriate operation (the number of elements contained in the group defines the order of the group). Often, a cyclic group of order $m$, $Z_m$, can be used to classify directional variants, and a cyclic group of order $n$, $Z_n$, to classify antiphases. A vertex at which $l$ domain boundaries merge can be treated as a topological vortex $Z_l$, with vorticity in real space or topological pseudo-vortex defined in the domain-boundary energy diagram, which depicts the low-energy domain boundaries. We will not distinguish pseudo-vortex from vortex for the sake of simplicity.

In this Review, we introduce a new scheme of $Z_m \times Z_n$ domains and $Z_l$ vortices to classify various vortex-domain configurations in a wide range of complex materials such as multiferroics[14,34,39,48,49], improper ferroelectrics[44,50-55], layered transition-metal dichalcogenides[56-59], bilayer graphene[60], and magnetic superconductors[61]. Vortex-like domains due to surface (or charge) boundary conditions or spatial confinement constitute 'extrinsic' domains (sometimes they are also called closure or circular domains) and were reported in various nanoscale particles or heterostructures[62-66]. This Review focuses on 'intrinsic' vortex domains, which usually form through thermodynamic processes or phase transitions. Nontrivial $Z_m \times Z_n$ domain topologies can exist in domain configurations with Ising-type order parameters (such as positive and negative polarity or left and right chirality) or with 2D order parameters (such as vectors in a plane). A configuration of large-range domains and domain boundaries can exhibit an aperiodic order with the topology defined by $Z_m \times Z_n$ domains with $Z_l$ vortices, analogous to the aperiodic order in the Fibonacci sequence, in the Penrose tiling and in quasicrystals[67,68]. These aperiodic orders are certainly different from periodic orders in real space, and are accompanied by nontrivial mathematical and topological rules. These rules have a similar role to that of self-similarity, which ensures that similar shapes develop at all length scales in the formation of fractals and dendrites. We highlight experimental observations of various domain topologies in functional materials and discuss the topological protection of domain configurations, switching kinetics, domain boundary energy diagrams and of the relevant physical properties of domain boundaries. The recent progress in the understanding of domain topologies is far-reaching, as it is relevant not only for the field of materials science, but also for mathematics (graph theory)[54,56], statistics (network theory)[69], and even cosmology (such as in the formation of cosmic strings)[53,70,71]. We also revisit some well-known materials in view of the new classification and conclude by overviewing the perspectives in domain topology engineering.

**$Z_m \times Z_n$ domains and $Z_l$ vortices**

At the beginning of the 20th century, the concept of domain structures was introduced by Weiss based on the idea of regions with different magnetization, which has been at the basis of all later theories[17,72,73]. The concept of domains inspired the exploration of magnetic domain images; for example, using the Bitter patterns studied by Bitter[73] and by von Hámos and Thiessen[74] in 1931. Later on, ferroelectric domains have been studied in materials such as potassium dihydrogen phosphate[75,76] in 1944, and Rochelle salt[17] and barium titanate[77,78] in 1948. A generalized definition of ferroic domains was later given by Aizu[16], who described them as domains that have "two or more orientation states in the absence of magnetic field, electric field and mechanical stress and can shift from one to another of these states by means of a magnetic field, an electric field, a mechanical stress, or a combination of these." In this Review, $Z_m$ denotes $m$ cyclic directional variant domains, which are associated with Aizu's orientation states[16]. Directional variants are related to suppressed or broken symmetry operations in a low-symmetry state that forms from a high-symmetry parent state through the ordering of charge, spin, orbital or chiral degrees of freedom. Order parameters associated with directional variants can be, for example, 3D vectors[17]. In this Review we focus on Ising-type and 2D discrete *xy*-type directional variants. Note that directional variants are related to each other through a symmetry operation such as a rotation, mirror reflection, space inversion, rotoinversion or time-reversal symmetry, other than through a translational operation (Table 2). Translational domains are identical in terms of the orientation of their tensor properties, but have different translational symmetry.

We first give a pedagogic example of an antiferromagnetic (AFM) spin chain with two choices of AFM spin orientation (↑↓, up-down, or ←→, left-right), corresponding to two directional variants. The interface between the two directional variants is a directional variant boundary (DVB) (FIG. 1a). Translational variant domains are related to phase shifts in translation symmetry operations, and their boundaries are often called antiphase boundaries (APBs); for example, there is a π phase shift at the translation variant boundary of ↑↓ ↓↑ ↑↓ (FIG. 1a). An analogous structural APB can be associated with a fraction of a lattice translation vector, but translational variant domains are identical in terms of atomic arrangements and tensor properties within each domain. The term $Z_n$ denotes *n* cyclic antiphase domains, which were first reported in the work of Johansson and Linde on the diffraction patterns of an AuCu alloy in 1936 [79-81].

Historically, the study of directional variant domains ($Z_m$) and of antiphase domains ($Z_n$) have developed independently, even though the basic concepts are closely related and the two types of domains have many aspects in common; for example, they often arise as the result of a phase transition. Directional variant domains are often associated with macroscopic physical properties such as magnetization, ferroelectric polarization and structural chirality, whereas antiphase domains arise from translational phase shifts of local distortions or from atomic arrangements and, as a consequence, tend to be degenerate in terms of macroscopic properties. To characterize the connectivity of $Z_m \times Z_n$ domains it is useful to consider the domain vertices at which several variant domains meet.

For each vertex, the vorticity can be defined in the $Z_m \times Z_n$ phase space, which is discussed in detail in the following section. A vertex at which $l$ variant domains meet is called a $Z_l$ vortex[82]. Obviously, $l$ domain boundaries also meet at a $Z_l$ vortex. The total angle around a vertex in the $Z_m \times Z_n$ phase space must be an integral multiple of $2\pi$ [82], and a topological charge or winding number $n$ is assigned when the relevant vectors rotate clockwise by $2\pi n$ around the vertex. For example, in a clockwise cycle, an increment in angle ($+2\pi$) is counted as a vortex and a decrement ($-2\pi$) as an antivortex. A domain topology with $Z_m \times Z_n$ domains and $Z_l$ vortices not only reflects the presence of low-energy domain boundaries, which can be represented by a domain-boundary energy diagram, but also implies a coherent continuity of atomic matching across domain boundaries. Another physical implication of the $Z_l$-vortex topology is that the creation of a vortex core is naturally accompanied by the simultaneous appearance of $l$ incident domain boundaries, which are very different from, for example, parallel stripe-like boundaries. Therefore, the energy barrier of the domain boundaries is also the energy barrier that guarantees the stability of an existing $Z_l$ vortex. We will further discuss the topological nature of $Z_m \times Z_n$ domains and $Z_l$ vortices in the following sections.

The simultaneous appearance of directional variant domains and antiphase domains opens new grounds for materials research, in particular for functional materials with strong interactions or with structural boundaries with ferroic order and interlocking nature[14,18,19,83]. Before discussing real materials, we consider AFM domains with possible $Z_3$ vortices. Figure 1a shows a $Z_3$ vortex and an antivortex consisting of two directional variant domains (↑↓ and →←) and one antiphase domain (←→) (which is antiphase with respect to the domain →←). Counting along the clockwise direction, the spins exhibit a $+2\pi$ rotation (vortex, green arrow), whereas in the counter clockwise direction the spins show a $-2\pi$ rotation (antivortex, purple arrow). We name the resulting domain structure as $Z_2 \times Z_2$ domains with $Z_3$ vortices, where the first $Z_2$ stands for two directional variant domains and the second $Z_2$ indicates two antiphase domains. The classification of $Z_m \times Z_n$ links all possible domain states to the order parameters and relevant bulk symmetry of a material, whereas the $Z_l$ vortex is a consequence of the energetics of domain boundaries and vortices. Thus, $l$ is not necessarily equal to the total number of domain states ($m*n$) and a simple comparison of $l$ and $m*n$ already reveals the basic energetics of the domain boundaries and vortices.

**Dielectrics — 4-fold vortices**

One intriguing structural example of $Z_2 \times Z_2$ domains with $Z_4$ vortices was recently found in a Ruddlesden-Popper-series bi-layered perovskite (RP327), $Ca_2Sr_1Ti_2O_7$ (FIG. 1b)[50]. Oxygen octahedral distortions in (layered) perovskites often have a crucial role in determining the functional properties of these materials[84-88]. For example, orthorhombic $Ca_3Ti_2O_7$ ($o$ state, space group $A2_1am$) exhibits simultaneously oxygen octahedral tilts (diagonal with respect to the in-plane tetragonal directions) and rotations, and reaches a switchable bulk polarization[44,89,90] of ~8 $\mu C \cdot cm^{-2}$. Conversely, $Sr_3Ti_2O_7$ forms in a tetragonal ($t$ state, $I4/mmm$) structure with undistorted oxygen octahedra[91]. Interestingly, a new tetragonal structure ($t'$ state, $P4_2/mnm$)[50] can be stabilized in $Ca_2SrTi_2O_7$; in this structure oxygen octahedral tilts ($a^-a^0c^0$ in the Glazer notation) without any rotations develop along the in-plane tetragonal directions (the $<100>_t$ directions, FIG. 1c). Because of the underlying square lattice, 4 symmetry-equivalent domains — $Z_1 \times Z_4$ domains — may exist; four translational variants are associated with the translation vectors (1/2, 1/2, 0), (0, 1/2, 1/2) and (1/2, 0, 1/2), and out-of-plane phase shifts are associated with the latter two translation vectors. However, the domain topology can be also classified as $Z_2 \times Z_2$ domains if the in-plane order parameters (octahedral tilts) of a single bilayer are considered, as there are two directional variants ([010]$_t$–tilts producing the domains 1 and 3 and [100]$_t$–tilts producing the domains 2 and 4 in FIG. 1b) and two translational variants (associated with the translation vector (1/2, 1/2, 0))[50].

Most strikingly, dark-field TEM images (FIG. 1d) reveal $Z_4$ vortices of eight-state: four bright-contrast domains form a vortex-like pattern (a $Z_4$ vortex) with 90°–rotating apical oxygen distortions (FIG. 1d, red arrows) and four dark-contrast domain boundaries at which apical oxygen displacements are along the $<110>_t$ directions (FIG. 1c-d, blue arrows). The octahedral tilting angles change by 45° consecutively around the vortex core as a consequence of the active $X_3^-$ mode, which is related to oxygen tilting[50]. If tilting vectors are given for one domain and one domain boundary, all domains, domain boundaries, vortices and antivortices can be identified. In this sense, the $Z_2 \times Z_2$ domains with $Z_4$ vortices are topologically protected. The topological defects observed in FIG. 1d can be identified as a type (i) and a type (ii) vortex, and type (iii) and type (iv) antivortex. Vectors in a type (i) vortex rotate in the opposite sense compared to those in a type (ii) vortex, but they have the same topological charge (+1), thus both of them are vortices. A $Z_4$ vortex is always surrounded by antivortices and vice versa; for example, in FIG. 1d the type (i) vortex is connected to the type (iii) and type (iv) antivortexes. The absence of $Z_3$ vortices, as in the simple AFM domain configuration discussed earlier, reveals the absence of (1/2, 1/2, 0)-type translational variant boundaries, such as the APB between domains 1 and 3, which, in turn, indicates that APBs have a high energy cost. The observation of $Z_2 \times Z_2$ domains with $Z_4$ vortices resulting from the presence of eight degrees of freedom in two dimensions provides an example of a unique real-space topology in which domains and domain boundaries are intricately intertwined[50].

**Ferroics — 6-fold vortices**

*Hexagonal R(Mn,Fe)O₃*

The schlieren textures observed in nematic or smectic liquid crystals, consisting of points from which 2 or 4 dark brushes depart, are arguably the most celebrated example of topological defects with the cloverleaf pattern[27]. $h$-RMnO$_3$ (R=Sc, In, Y, Dy–Lu) with improper ferroelectricity is the first 'hard' condensed matter system exhibiting cloverleaf patterns of $Z_2 \times Z_3$ domains and $Z_6$ vortices (FIG. 2a). The onset temperatures for ferroelectricity in $h$-RMnO$_3$ are around 1,250 K, 1,403 K, 1,523 K, and 1,672 K for YMnO$_3$, ErMnO$_3$, TmMnO$_3$, and LuMnO$_3$, respectively[53,71,92]. The structural phase transition leading to ferroelectricity can be described by a Mexican-hat-type potential (FIG. 2b)[42,70], and is manifested by a trimerizing distortion — the $K_3$ mode — in which three MnO$_5$ bipyramids tilt towards (or away from) their common oxygen atom. The $K_3$ mode is coupled to the secondary polar $\Gamma_2^-$ mode through the trimerization phase $\varphi = 30°*n$ (with $n$ even)[25,26,37,38,42,70,83]. For positive $\Gamma_2^-$, the coupling favours $K_3$ tilt angles of 0°, 120°, and 240° with up-polarization (+), whereas negative $\Gamma_2^-$ favours angles of 60°, 180°, and 300° with down-polarization (−). Directional variants $Z_2$ are associated with + and − ferroelectric polarization, whereas the unit-cell-trimerizing distortion in the *ab*-plane gives rise to three types of antiphase domains (α, β and γ, FIG. 2a), forming a √3×√3 $Z_3$ superstructure[14]. These six antiphase and ferroelectric domains cycle around a merging point with alternating polarization and trimerization antiphases in two different domain configurations: (α+, β-, γ+, α-, β+, γ-) and (α+, γ-, β+, α-, γ+, β-), which can be viewed as a vortex and an antivortex, respectively (a vortex is shown in FIG. 2a). Across each domain boundary, all the MnO$_5$ bipyramids rotate coherently by 60° (FIG. 2c), thus the total oxygen distortion around one vortex core is +2π. DVBs (ferroelectric domain boundaries) and APBs are mutually interlocked to form $Z_6$ vortices[14,38,42,93], and each domain is always surrounded by an even number of vortices and antivortices[54,69] (Box 1). *h*-RFeO$_3$, which exhibits improper ferroelectricity (and weak ferromagnetism, unlike *h*-RMnO$_3$, which is antiferromagnetic) displays a similar domain topology[49,94].

$Z_2×Z_3$ domains and $Z_6$ vortices in *h*-RMnO$_3$ have been probed using various techniques, such as transmission electron microscopy (TEM)[14,29-31,36,95-98], scanning secondary-electron microscopy[99], piezoresponse force microscopy[33,70,71,92,96,100], magnetic force microscopy[34], conductive atomic force microscopy[32,33], magnetoelectric force microscopy[35], optical microscopy[54,96], optical second-harmonic generation[92,101], X-ray photoemission electron microscopy[102] and scanning microwave impedance microscopy[103]. Details of the physical properties *h*-RMnO$_3$ can be found in several topical reviews; for example, REF. 15 for magnetoelectric coupling and REFS 48 and 104-106 for multiferroic domain structures. This section focuses on the recent progress in the understanding of domain topology and its implications, covering domain topology evolution[42,53-55,69,71,96] and scaling behaviours[53,70,71], the unfolding of $Z_6$ vortices[96,107], and the duality of topological vortices[83].

***Domain topology evolution and scaling behaviours.*** Onsager and Feynman[108] predicted that the restoration of a continuous $U(1)$ symmetry at the critical temperature $T_c$ marking a phase transition, for example that from the superfluid to the normal state of $^4$He [REF 40], can occur through the proliferation of topological defects such as vortex lines and loops[71,109]. Therefore, the condensation of topological defects significantly depends on whether the initial crystal growth or annealing temperature, $T_i$, is lower or higher than $T_c$, and also on the cooling rate across $T_c$[53,70,71,109]. If $T_i < T_c$, the fluctuating vortices present at $T_i$ tend to shrink and disappear during cooling. Conversely, if $T_i > T_c$, vortices span the whole system and disappear at a much slower rate, so that the network of condensed vortices can survive. Experimentally, this phenomenon is clearly observed: in LuMnO$_3$ ($T_c$=1,672 K), $Z_6$ vortices remain in the final state if $T_i$=1,698 K, $> T_c$ (FIG. 2d), but annular patterns or stripes are observed if $T_i$=1,633 K, $< T_c$ (FIG. 2e)[71]. These two distinct behaviours are observed even for $T_i$=1,673 K and $T_i$=1,671 K. If the cooling rate across $T_c$ is varied from 0.5 ºC/h to 300 ºC/h vortex-antivortex domain patterns persist, but the vortex density varies with a power law as a function of the cooling rate, which is consistent with the so-called Kibble–Zurek mechanism[40,55,70,71,109]. In addition, the vortex networks can be of two different types: type I vortex domains have roughly equal fractions of opposite polarization domains (FIG. 2f), type II vortex domains have one dominant type of polarization domains, surrounded by unfavoured-polarization domains that can be as narrow as 8 unit cells (FIG. 2g)[31,54]. The switching between type I and type II vortex domains occurs in the presence of external electric fields[31,96] or through an internal self-poling effect derived from oxygen off-stoichiometry[96]. The extensive statistical analysis of type I vortex domains reveals a log–normal statistical distribution, whereas the statistical distribution of type II vortex domains fits a scale-free power-law distribution. The different statistical distributions uncover the 'preferential attachment' mechanism underlying the evolution from type I to type II vortex domains[69], for which $N$-gons with larger $N$ (Box 1) have a higher probability to coalesce with other segments.

***Unfolding of $Z_6$ vortex cores.*** As discussed above, the stripe-type or annular domain topology is stable when $T_i < T_c$. The control of vortices, such as the transformation of vortices into stripes, has not been achieved with external electric fields, probably because of the tendency toward charge cancellation at vortex cores[31]. Instead, shear strain with a strain gradient applied at high temperatures has been demonstrated to be an effective tool for vortex control[107]. Shear strain induces a Magnus-type force[110] acting in opposite ways on vortices and antivortices along the direction normal to the shear strain. In addition, the strain gradient results in a force pulling vortex-antivortex pairs along the strain gradient direction (FIG 2h). All vortices remain at the transformation boundary, but antivortices are expelled. The result is that vortex domains are converted to stripe domains, which is consistent with a 'Φ-staircase' state with a nonzero average gradient of the phase Φ (Φ is the azimuthal angle describing the displacement of apical oxygens)[42]. This Φ-staircase-like state consists in topological stripes with monochirality, corresponding to a fixed sequence of the six trimerized phases that are either vortex-like (α+, β-…) or antivortex-like (α+, γ-…). The underlying physics is that of topological stability: an individual topological vortex is rather stable against small perturbations[31,96], but can be annihilated or expelled in vortex/antivortex pairs, in close analogy with Dirac's particles and antiparticles.

*Duality of topological vortices.* By symmetry, the condensation of the $K_3$ mode allows three sets of $K_3$ tilting angles ($\varphi=30°*n$), corresponding to three phases with distinct space groups. The known $Z_6$ ferroelectric vortex patterns correspond to even $n$ with a polar $P6_3cm$ space group (FIG. 2i). Odd values of $n$ correspond to partially undistorted antipolar (PUA) states with non-polar $P\bar{3}c1$ symmetry, whereas an intermediate tilting angle of the $MnO_5$ bipyramid (non-integer $n$) results in polar $P3c1$ symmetry[52,83,111]. In the ferroelectric Mexican-hat-like potential landscape discussed earlier (FIG. 2b), there are six minima located at $30°*n$ with $n$ even, corresponding to the six ferroelectric domains; the energy barriers separating them lie at the intermediate tilt angles of $30°*n$ with $n$ odd, which are, in fact, in the PUA state. In the PUA potential landscape[83] the energy minima are found at angles of $30°*n$ with $n$ odd, complementary to those of the ferroelectric case. Thus, the PUA state may exhibit $Z_2 \times Z_3$ PUA domains with two oxygen-rotation directions plus three different phase shifts (τ, υ, μ, FIG. 2i). Experimentally, the ground state of $h$-InMnO$_3$ is in the ferroelectric $P6_3cm$ state, but compressive chemical strain applied through gallium doping can stabilize the PUA ground state (FIG. 2j). The $Z_2{\times}Z_3$ PUA domains in $h$-In(Mn,Ga)O$_3$[83,111] consist of six domains with three 'hidden' domain boundaries alternating with three readily visible domain boundaries. The hidden domain boundaries, which are associated with low strain fields, exhibit a progressive evolution of In-ion displacements and polyhedral tilt angles, corresponding to intermediate $P3c1$ and ferroelectric symmetry[83]. $Z_6$ PUA vortices and $Z_6$ ferroelectric vortices are in a similar topological configuration, but are 'dual' to each other, in the sense that the domain boundaries of $Z_6$ PUA vortices are ferroelectric and vice versa. The switching between these dual topological objects through thermal variations or external perturbations such as electrics fields, as well as its kinetics, needs to be further investigated.

## *Layered di-chalcogenides: 2H-Fe$_{1/3}$TaS$_2$*

Layered transition metal di-chalcogenides, consisting of MC$_2$ layers (M: transition metal elements, C: chalcogen element) have been extensively investigated because they host several interesting physical phenomena, such as charge-density waves[112-114], superconductivity[113,115], few-layers field effect[116-118] and Weyl semi-metallic behaviour[119,120]. Ion intercalation between the MC$_2$ layers can alter their superstructures and physical properties[121,122]. For example, Fe-intercalated TaS$_2$, 2H-Fe$_{1/3}$TaS$_2$, (the 2H denotes a trigonal prismatic coordination), which is ferromagnetic, shows a domain topology with $Z_2{\times}Z_3$ domains and $Z_6$ vortices (FIG. 3a,b)[56] identical to that of ferroelectric $h$-RMn(Fe)O$_3$. In the domain topology of 2H-Fe$_{1/3}$TaS$_2$, $Z_3$ represents three antiphase domains in a given $\sqrt{3}{\times}\sqrt{3}$ superstructure originating from the ordering of the Fe ions, and the directional variant $Z_2$ is associated with two types of structural chirality in the corresponding chiral $P6_322$ space group[123]. Across each domain boundary, all the sulfur distortions in one of two adjacent layers rotate coherently by 120º (FIG. 3a), thus the total sulfur distortion around one vortex core is +4π. Dark-field TEM images taken under Friedel's-pair-breaking conditions[56,124] clearly show chiral domains without centrosymmetry (FIG. 3b). The interlocking of DVBs (which are chiral domain boundaries) and APBs leads to topologically protected $Z_6$ vortices.

The typical domain size of $Z_2 \times Z_3$ domains with $Z_6$ vortices in 2H-Fe$_{1/3}$TaS$_2$ is ~3 µm, which is around 20 times larger than that of Fe-intercalated compounds with lower Fe content, as discussed later. A relatively low density of topological defects in 2H-Fe$_{1/3}$TaS$_2$ results in a weak pinning of magnetic domain boundaries and in a small magnetic coercivity (FIG. 3c)[56]. Most likely, 2H-Fe$_{1/3}$TaS$_2$ is not the only transition metal di-chalcogenide to display a domain topology; for example, 2H-M$_{1/3}$NbS$_2$ (M=Cr, V, Co, Ni, Mn and Fe)[123,125] and 2H-M'$_{1/3}$TaS$_2$ (M'=Ni, Pb, and Co)[122,123] belong to the same chiral symmetry group, and are expected to exhibit a similar domain topology. This findings in layered di-chalcogenides with ferromagnetic metallicity[122,126] demonstrate the universality of the domain topology that we are discussing and provide new insight in the relationship between bulk properties and mesoscopic domain topology.

**Improper ferroelectrics — 3-fold vortices**

The condition of $l=m*n$ is satisfied for $Z_2 \times Z_3$ domains with $Z_6$ vortices, as well as for $Z_2 \times Z_2$ domains with $Z_4$ vortices. However, $Z_m \times Z_n$ domains with $Z_l$ vortices with $l<m*n$ can be realized, especially for sufficiently large $m*n$ (FIG. 4). A representative material is RP327 $o$-Ca$_{3-x}$Sr$_x$Ti$_2$O$_7$ ($x \leq 0.9$), which exhibits $Z_4 \times Z_2$ domains with $Z_3$ vortices (FIG. 4a). Bulk single crystals of $o$-Ca$_{3-x}$Sr$_x$Ti$_2$O$_7$ were the first material in which hybrid improper ferroelectricity[43,87,89,90,127-129] was experimentally demonstrated in a bulk sample[44]. The simultaneous presence of cooperative octahedral tilts (order parameter X$_3^-$, Glazer notation $a^-a^-a^0$) and octahedral rotations (X$_2^+$, $a^0a^0c^+$) results in a polar $A2_1am$ space group with 4 directional variants (FIG. 4e), as revealed by in-plane piezoresponse force microscopy (FIG. 4c)[44]. Two possible choices of the origin of a $\sqrt{2} \times \sqrt{2}$ superstructure result in additional $Z_2$ antiphase domains with the translational vector (1/2, 1/2, 0), which corresponds to a simultaneous sign change of the X$_3^-$ and X$_2^+$ order parameters[43,89,130] (FIG. 4e). Dark-field TEM images (FIG. 4b) show that each of four polarization-direction states is degenerate with two antiphase domains, and these eight structural variants form a $Z_4 \times Z_2$ domain structure with $Z_3$ vortices and five distinct types of domain boundaries[43]. Note that this domain topology is directly relevant to the presence of abundant charged domain boundaries, which are, in general, energetically unfavourable (FIG. 4c and 4d)[44,131]. Unlike the formation of antiparallel domains separated by charge-neutral domain boundaries, typical of 180º ferroelectric domain boundaries[1], the appearance of vortex-connected domain configurations indicates a negligible energy difference among different types of domain boundaries, such as charged and non-charged boundaries. The relevant domain-boundary energy diagram, representing all possible low-energy domain boundaries, is a hyper-tetrahedron in 7 dimensions; each of all eight states, represented by the vertices of the hyper-tetrahedron, is connected to all other states and the connecting edges, representing domain boundaries, consist of four types of DVBs and one APB (FIG. 4f). Eight degenerate states or domains are labelled 1±, 2±, 3±, and 4±, denoting the octahedral distortions in which the $X_3^-$ mode adopts one of the four tilts (toward one of the quadrants labelled 1 to 4) accompanied by + or - rotations of the $X_2^+$ mode (resulting in a combined distortion pattern of polar $a^-a^-c^+$, FIG. 4e). The three edges of each triangular face of the hyper-tetrahedron energy diagram correspond to a $Z_3$ vortex or antivortex, depending on the cyclic order of the domain boundary arrangement. Note that if, for example, the APB is associated with an energy much higher than that of the other domain boundaries, the APB is fully avoided, and $Z_l$-type vortex defects with $l>3$ can occur. However, experimentally only $Z_3$-type vortices have been observed. In terms of the vortex density, the size of $Z_3$-vortex domains varies from ~0.5 to ~20 µm with Sr doping[43,44], but it changes little in response to heat treatments, which indicates that the formation of $Z_4 \times Z_2$ domains with $Z_3$ vortices is associated with a 1st-order phase transition. This is very different from what happens, for example, in $h$-RMnO$_3$, which shows a strong dependence of the vortex density on the sample cooling rate across a 2nd-order phase transition temperature[71].

One of the most interesting aspects of the domain topology approach is the pairwise nucleation/annihilation observed during polarization switching[43]. In-situ poling of $o$-Ca$_{3-x}$Sr$_x$Ti$_2$O$_7$ ($x=0$ and 0.45) using a dark-field TEM technique unveils intriguing domain switching kinetics, which can be understood in terms of a zipper-like role of APBs: APBs act as reversible creation/annihilation centres of two different DVBs pairs (and also of $Z_3$ vortex–antivortex pairs) in 90º or 180º polarization switching. A schematic illustration of 180º ferroelectric polarization switching through the splitting or coalescence of an APB and two $FE_t$ + $FE_r$ DVBs ($FE_t$ and $FE_r$ boundaries stand for ($X_3^-$)-type tilting and ($X_2^+$)-type rotation DVBs, respectively), accompanying a $Z_3$ vortex–antivortex pair creation/annihilation, is shown in FIG. 4g. The splitting accompanies the creation of a vortex–antivortex pair, the coalescence its annihilation. Thus, the topological concept of $Z_4 \times Z_2$ domains with $Z_3$ vortices is useful to understand the formation and kinetics of domains and domain boundaries of hybrid improper ferroelectrics. A similar domain topology is expected in several other compounds; for example, $Z_4$-type in-plane directional variants, accompanying $Z_2$-type antiphases associated with orthorhombic lattice distortions, commonly exist in layered compounds such as the Dion-Jacobson phases (A'[$A_{n-1}B_nO_{3n+1}$])[132-134] and the Aurivillius phases ([$Bi_2O_2$]$^{2+}$[$A_{n-1}B_nO_{3n+1}$]$^{2-}$) [87,135].

## (Anti)-ferromagnets — 3-fold vortices

The presence of four types of domains and three types of domain boundaries with $Z_3$ vortices is unique in the sense that in this case $l$ is equal to the total number of domain boundary types. Two examples are the antiphase domains related to the ordering of the iron atoms in $Fe_{1/4}TaS_2$[56] and the plaquette antiferromagnetic domains in $Sr_2VO_3FeAs$, an iron-based superconductor[61].

### *Layered di-chalcogenides: 2H-Fe$_{1/4}$TaS$_2$*

The ordering of the intercalated iron ions in $Fe_{1/4}TaS_2$ results in a 2×2 superstructure with four different crystallographic origins, A(0, 0), B(1/2, 0), C(1/2, 1/2) and D(0, 1/2)[56]. Four antiphase domains are identified as AA, BB, CC and DD in the presence of in-phase stacked 2D supercells (FIG. 3d), and three types of APBs with relative phase shifts of (π, 0), (0, π) and (π, π) can be identified. The relevant domain-boundary energy diagram of $Z_1 \times Z_4$ domains and $Z_3$ vortices is a 3D tetrahedron[136] (FIG. 3c), in which each vertex is connected to all other vertices through 3 edges that correspond to the three types of APBs. Identifying a couple of domains in the $Z_1 \times Z_4$ domain pattern is not sufficient to identify all the remaining domains (FIG. 3d), because the pattern is not topologically protected, as discussed later. However, because three types of APBs always merge at each $Z_3$ vortex core, we can define a vortex (antivortex) with topological charge +1 (-1) for a merging point at which three types of APBs order in a clockwise (counter-clockwise) sequence in the domain-boundary energy diagram (as indicated by the green arrow in FIG. 3c for a vortex). At this point, topological charge conservation works; in other words, the creation/annihilation of vortex–antivortex pairs changes the domain topology.

### *Iron-based superconductor: $Sr_2VO_3FeAs$*

The same topology with $Z_1 \times Z_4$ domains with $Z_3$ vortices is observed in $Sr_2VO_3FeAs$, in which a plaquette AFM order appears to coexist with superconductivity[137,138]. The plaquette AFM order is thought to be stabilized by quantum fluctuations[61,137-140], and its presence was first demonstrated using spin-polarized scanning tunnelling microscopy. Four origins for the AFM iron spin plaquette lead to four AFM antiphase domains (FIG. 3f). Because the spin directions are not identified, possibly due to spin fluctuations or to the presence of a single directional variant, we may assign $Z_1 \times Z_4$ domains to the plaquette AFM order; in fact, the plaquette AFM domain configurations observed by spin-polarized scanning tunneling microscopy are consistent with $Z_1 \times Z_4$ domains with $Z_3$ vortices (FIG. 3g)[61]. Similar with what happens in $Fe_{1/4}TaS_2$, three types of APBs always merge at one point, thus topological charges can be defined in terms of domain boundary configuration. One example of topological charge conservation is shown in FIG. 3g: one vortex–antivortex pair annihilates into (or is created from) topological 'vacuum' with zero net topological charge. $Z_1 \times Z_4$ domains and $Z_3$ vortices certainly require further investigations, especially to understand the formation, manipulation, and kinetics of domain boundaries.

**Other examples**

*Charged-density waves: 1T-TaS$_2$*

The concept of $Z_m \times Z_n$ and $Z_l$ vortices can be extended to understand the domains of a metastable state induced by an external electric field at a local scale[57,58]. 1T-TaS$_2$ (where 1T denotes an octahedral coordination) is known to form a charged-density wave (CDW) state in which the in-plane Ta lattice is susceptible to a David star-type ($\sqrt{13} \times \sqrt{13}$) deformation, accompanying an electronic reconstruction to a Mott insulating state[112,118,141]. An electric voltage pulse from a scanning tunnelling microscope creates a textured CDW domain pattern in 1T-TaS$_2$ with a majority of triple junctions[57,58] (FIG. 5a). In the David-star-type CDW in 1T-TaS$_2$, *1×13*-state domains with $Z_3$ vortices are associated with 13 antiphase domains (with no directional variants)[57,58]. Note that no cyclic relationship exists among the 13 antiphase domains in 1T-TaS$_2$. A specific translational vector –*a*+*b* (where *a* and *b* are primitive hexagonal unit vectors) is preferred at antiphase boundaries, thus each antiphase domain always has six low-energy neighbouring domains instead of twelve[57,58]. The translational vector is defined by the relative shift of the David-star-pattern between two neighbouring domains[58]. Thus, the relevant domain-boundary energy diagram is the so-called 'regular graph of 13 vertices with degree 6' (http://www.mathe2.uni-bayreuth.de/markus/reggraphs.html) (FIG. 5b). A $Z_3$ (anti)vortex corresponds to a loop connecting three vertices in a (counter-)clockwise cycle in the domain-boundary energy diagram. For example, a loop of 0–7–2 corresponds to a $Z_3$ vortex, whereas a loop of 0–2–7 corresponds to a $Z_3$ antivortex. It should be emphasized that this energy diagram is very different from the hyper-tetrahedron discussed above, in which each vertex (phase) is always connected to all others. Large-scale images suitable for a complete topological analysis of this system are needed. Our classification and topological considerations will be particularly useful for the understanding of the domain configurations of layered materials with large super cells[142-144].

*Charge/orbital order: Pr(Sr$_{0.1}$Ca$_{0.9}$)$_2$Mn$_2$O$_7$*

Another interesting topic is the understanding of the evolution of domain topology in systems that undergo a series of ferroic phase transitions. For example, RP327 *o*-Pr(Sr$_{0.1}$Ca$_{0.9}$)$_2$Mn$_2$O$_7$ (PSCMO) exhibits two distinct charge- and orbital-ordered phases with a hysteretic transition near room temperature (T$_{CO1}$: ~330 K and T$_{CO2}$: ~295 K)[51,145]. The CO1 state is antiferroelectric with orbital stripes running along the *a*-axis, whereas the CO2 state is ferroelectric with orbital stripes running along the *b*-axis (FIG. 5c)[51,145,146]. Both the CO1 and CO2 states are supposed to exhibit $Z_4 \times Z_2$ domains with four directional variants (antiferroelectric for CO1 and ferroelectric for CO2) and two antiphase domains originating from a cell doubling[145]. Scanning microwave impedance microscopy (sMIM), a scanning probe technique that measures GHz-range local electromagnetic response, has provided evidence of enhanced conducting charged domain boundaries in both states[51]. In the sMIM domain patterns, $Z_4$ vortices are dominant within the spatial resolution, but $Z_3$ vortices also exist. In any case, antiphase domains are not observable in sMIM, so the exact meaning of sMIM domain patterns still needs to be clarified. Nevertheless, if antiphase domain boundaries are identified, PSCMO will be an ideal system to explore the evolution of domain topology through antiferroelectric to ferroelectric phase transitions and the relationship between domain topology and conducting domain boundaries.

**Implications of vortex domains**

*Topological protection*

When *l*=*m*\**n*, a topological protection of the domain configurations seems to exist, in the sense that a domain configuration changes only through the creation or annihilation of vortex–antivortex pairs and numerous domains are uniquely identified if two adjacent domains are identified. This topological protection naturally accompanies topological charge conservation. Topological protection and topological charge conservation work well in the case of $Z_2 \times Z_3$ ferroelectric domains with $Z_6$ ferroelectric vortices in *h*-R(Mn,Fe)O$_3$, $Z_2 \times Z_3$ PUA domains with $Z_6$ PUA vortices in *h*-In(Mn,Ga)O$_3$, $Z_2 \times Z_3$ chiral domains with $Z_6$ vortices in 2H-Fe$_{1/3}$TaS$_2$, and $Z_2 \times Z_2$ domains with $Z_4$ vortices in $t'$-Ca$_2$SrTiO$_7$. If $l<m*n$, such as for $Z_4 \times Z_2$ domains with $Z_3$ vortices in $o$-(Ca,Sr)$_3$Ti$_2$O$_7$ and $Z_1 \times Z_4$ magnetic domains with $Z_3$ vortices in Sr$_2$VO$_3$FeAs and 2H-Fe$_{1/4}$TaS$_2$, there is no topological protection of domain configurations. In other words, identifying a couple of domains is not sufficient to identify all domains, and the vortex–antivortex pair creation/annihilation is not the only way to change the domain configurations. However, as discussed earlier, in the case of $Z_1 \times Z_4$ magnetic domains with $Z_3$ vortices in Sr$_2$VO$_3$FeAs and 2H-Fe$_{1/4}$TaS$_2$, the relevant energy diagram is a 3D tetrahedron with 4 types of domains and 3 types of domain boundaries. In this particular case, the domain boundaries configuration, rather than the domain configuration, constitutes a $Z_3$ vortex–antivortex network. Furthermore, even though topological protection and topological charge conservation are not guaranteed, the zipper-like nucleation/disappearance of a new domain can accompany the creation/annihilation of $Z_3$ vortex–antivortex pairs near an APB in $o$-(Ca,Sr)$_3$Ti$_2$O$_7$.

*P-state clock and Potts models*

The formation of the domains/domain boundary configurations that we have discussed can be approximately described with either the clock[147-149] or Potts model[136,150-155] if the relevant phase transitions contain order–disorder-type characteristics, rather than having a pure displacement-type nature. In the $p$-state clock model, the energy ($E$) of a $Z_m \times Z_n$ domain configuration is $E = -J \Sigma S_i \cdot S_j = -J \Sigma \cos(\theta_i - \theta_j)$, where $\theta_i = 2\pi n_i/p$, $p = m*n$, $n_i$ varies between 1 and $p$[71,147] and $J$ is effective coupling constant between the 2D vectorial variables $S_i$ and $S_j$. The relevant energy diagram looks like a clock with $p$ ticks, which is the origin of the model name. The $p$-state clock model tends to result in $Z_l$ ($l = p = m*n$) vortices, corresponding to one turn around the clock. In the Potts model, the energy of a $Z_m \times Z_n$ domain configuration is $E = 2J\Sigma(1/p - \delta(s_i, s_j))$ where $\delta$ is the Dirac delta function and $s_i$ varies between 1 and $p$[136,155]. The relevant energy diagram is a ($p$-1)-dimensional hyper-tetrahedron. The face of the hyper-tetrahedron is triangular for any dimension, which is responsible for the formation of $Z_3$ vortices in the $p$-state Potts model[136]. Note that the Potts model is relevant to systems in which all possible domain boundaries have identical or similar energy, whereas in the clock model only the domain boundaries of adjacent states on the clock have the lowest energy. Topological vortices and antivortices with 2D vectorial degrees of freedom are rigorously defined in terms of topological charges (or winding numbers) in real space. Examples include $h$-R(Mn,Fe)O$_3$, $h$-In(Mn,Ga)O$_3$, 2H-Fe$_{1/3}$TaS$_2$ and $t'$-Ca$_2$SrTi$_2$O$_7$. However, in this Review we also use the terms vortices and antivortices when a vorticity or the opposite vorticity can be defined in the domain-boundary energy diagrams: examples include Sr$_2$VO$_3$FeAs, 2H-Fe$_{1/4}$TaS$_2$ and $o$-(Ca,Sr)$_3$Ti$_2$O$_7$. Note that even though some vortices and antivortices in domain-boundary energy diagrams are not rigorously-defined topological vortices and antivortices in real space, they can be created or annihilated in a pair-wise manner. In the case of 1×13-state domains with Z$_3$ vortices in 1T-TaS$_2$ only six (rather than twelve) domain boundaries for each domain have low energy, thus the relevant domain-boundary energy diagram is a regular graph of 13 vertices with degree 6. An extended Potts model, in-between the 13-state clock and the Potts model, is appropriate for this energy diagram.

All the materials discussed in this Review are listed in Table 2. They are mostly quasi-2D (that is, layered), and the 2D images of the domains and domain boundaries at their surfaces have been extensively discussed. However, the relevant interactions in bulk materials are evidently always 3D in nature. It has been argued that a 3D clock model is always associated with a continuous 2$^{nd}$-order phase transition, whereas a 3D Potts model accompanies a 1$^{st}$-order phase transition unless $p$=2 [136,150,154,156]. Note that a 3D $p$=2 (or 3) clock model is identical to a 3D $p$=2 (or 3) Potts model, and that $p$=2 corresponds to the Ising order parameter[148]. In this sense, the domain configurations of $h$-R(Mn,Fe)O$_3$, $h$-In(Mn,Ga)O$_3$, 2H-Fe$_{1/3}$TaS$_2$ and $t'$-Ca$_2$SrTi$_2$O$_7$ are associated with 2$^{nd}$-order phase transitions and are topologically protected. Conversely, Sr$_2$VO$_3$FeAs and $o$-(Ca,Sr)$_3$Ti$_2$O$_7$ are associated with 1$^{st}$-order phase transitions, and their domain configurations are not topologically protected.

Another interesting aspect is related to the Kibble-Zurek mechanism[40,109]. The topological defect density associated with a 2$^{nd}$-order transition is determined by the sample cooling rate across the phase transition temperature in such a way that slower cooling induces a lower density of topological defects. However, the topological defect density associated with a 1$^{st}$-order transition is not determined by the sample cooling rate across the phase transition temperature. Instead, it is determined by disorder, such as chemical impurities and lattice imperfections, which in this case is the most relevant factor for domain nucleation. Thus, the topological defect density of the domain configurations in $h$-R(Mn,Fe)O$_3$, $h$-In(Mn,Ga)O$_3$, 2H-Fe$_{1/3}$TaS$_2$ and $t'$-Ca$_2$SrTi$_2$O$_7$ is dominated by the sample cooling rate across the phase transition temperatures, whereas that of Sr$_2$VO$_3$FeAs, 2H-Fe$_{1/4}$TaS$_2$ and $o$-(Ca,Sr)$_3$Ti$_2$O$_7$ is mostly determined by the degree of chemical and crystallographic disorder.

**Old domains requiring new attention**

The so-called Dauphine twins in trigonal α–quartz[157,158] consist of two types of directional variants (related through a C$_2$ rotation) within a single chiral domain, and can often form 6-fold vertices at which three directional variants of each type merge (FIG. 5d)[157,158]. These vertices are not topologically protected, thus they can readily split, for example upon heating. It has been known for a long time that 2H-TaSe$_2$ [59] and γ-brass[159-161] can also have a domain configuration with similar 6-fold vertices, called discommensuration dislocations (DDs), and that these DDs are associated with one-dimensional phase shifts at domain boundaries (Box 2 shows various defects in 2D textures with 1D, 2D, and 3D vectorial order parameters)[5,59,147,162-168]. Bilayer graphene[60,169], which has two directional (inversion) variants and three antiphases, was found to exhibit similar 6-fold vertex domains (FIG. 5e). These vertices also seem topologically not protected, because they can split upon heating. It appears that the relevant classification is $Z_2 \times Z_3$ domains with $Z_3$ vortices, rather than $Z_2 \times Z_3$ domains with $Z_6$ vortices. Thus, the 6-state Potts model may be the relevant one in this case; there is no topological protection, and each 6-fold vertex can be, in fact, considered as four $Z_3$ vortices (FIG. 5f). Further experimental studies are required to understand the exact process of domain fragmentation and its relation with topological non-protection, and to elucidate how antiphase boundaries appear or disappear.

**Conclusions**

The classification of $Z_m \times Z_n$ domains and $Z_l$ vortices is universal and powerful, as it allows to understand macroscopic domain configurations in a wide range of materials with Ising-type or 2D order parameters. Additional topological considerations regarding domain configurations, such as topological protection, topological charge conservation and topological condensation, are particularly useful to understand the formation, evolution kinetics and statistics of domains and domain boundaries. New kinds of $Z_m \times Z_n$ domains and $Z_l$ vortices will be undoubtedly discovered. With further progress in imaging techniques with ultra-high spatial resolution, we can expect that this classification and the related topological considerations will be extensively employed and further fine-tuned. Even if the order parameters are three dimensional, the domain configurations of thin films and surfaces can be analysed in terms of the discussed classification and topological considerations. Extending this approach to truly 3D domain configurations is challenging. One area that still needs to be explored is the nano-scale electronic and optical spectroscopy of domain boundaries and vortices. The presented classification and the related topological considerations will provide a new platform for the engineering and manipulation of domains and domain boundaries to enable new functionalities.

**Acknowledgments**: we have benefited greatly by the discussion with Cristian D. Batista (U. of Tennessee), Dal Young Jeong (Soongsil U.) and Weida Wu (Rutgers U.). We thank Doohee Cho and Seong Joon Lim for critical reading of the manuscript, and Laura L. Cheong for providing schematics. This work is supported by the Gordon and Betty Moore Foundation's EPiQS Initiative through Grant GBMF4413 to the Rutgers Center for Emergent Materials.

**Box 1| Graph theory applied to domain configurations**

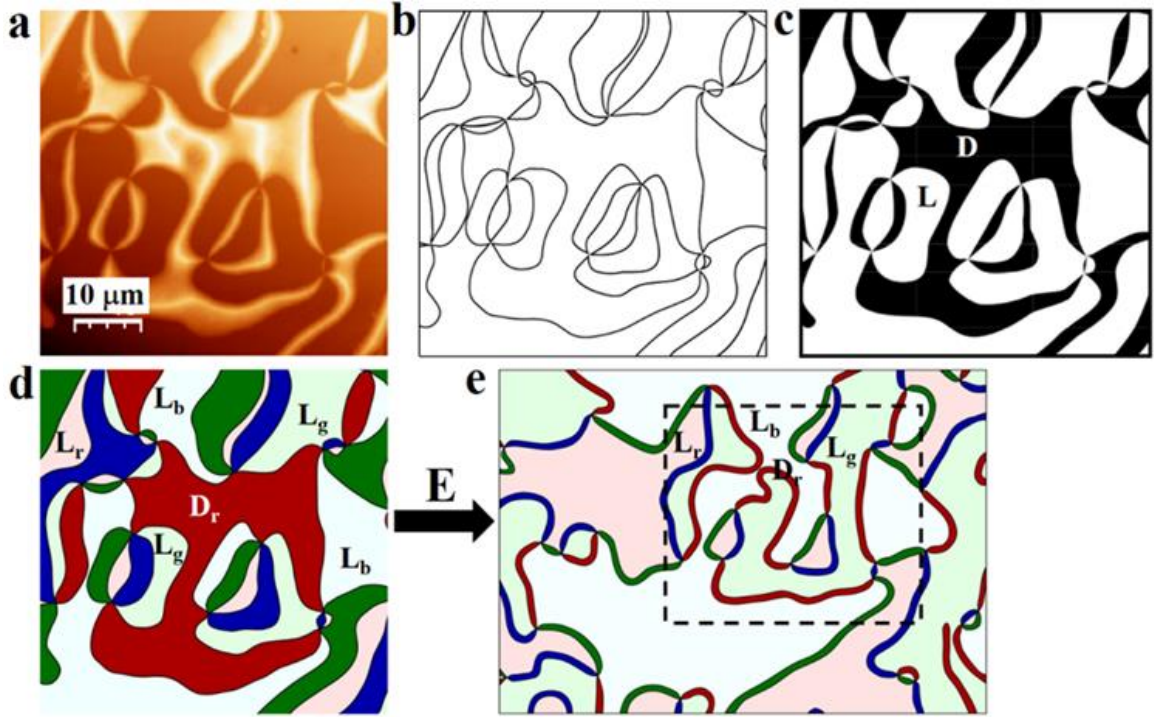

The four-color theorem in graph theory states that four colours are sufficient to identify all countries on a planar map in such a way that two bordering countries (not counting those that meet at a point) never have the same colour — this is called proper colouring. If all countries can be identified using three or fewer colours, then the map must have certain restrictions or rules.

In the language of graph theory, the domain boundaries of $Z_2 \times Z_3$ domains with $Z_6$ vortex/antivortex pairs in $h$-R(Mn,Fe)$O_3$, $h$-In(Mn,Ga)$O_3$, or 2H-Fe$_{1/3}$TaS$_2$ form a 6-valent graph with even $N$-gons. A 6-valent graph is a graph for which 6 edges merge at each graph vertex. A face of a graph loop with $N$ vertices on its boundary is called an $N$-gon. For example, the schematic of domain boundaries shown in panel **b** (which corresponds to the atomic force microscopy image of a chemically-etched $h$-RMnO$_3$ surface shown in panel **a**), is a 6-valent graph with even $N$- gons. All 6-valent graphs with even $N$-gons are 2-proper-colourable, meaning that no bordering domains have the same colour (panel **c**). However, a 6-valent graph with even $N$-gons can have a two-step tensorial proper colouring: after 2-proper colouring with dark (D) and light (L) (Panel **c**), a second-step proper colouring can be performed using blue (b), green (g) and red (r) in such a way that, for example, $D_r$ is surrounded only by $L_g$ and $L_b$, but never by $L_r$. (panel **d**). All six types of coloured domains have all kinds of even $N$-gons. However, after electric poling, all three types of D domains only have 2-gons, whereas all three types of L domains still have all kinds of even $N$-gons (panel **e**). Thus, electric poling is associated with a topological condensation of D domains to 2-gons, accompanying $Z_2$ symmetry breaking.

**Box 2| Defects in two-dimensional textures**

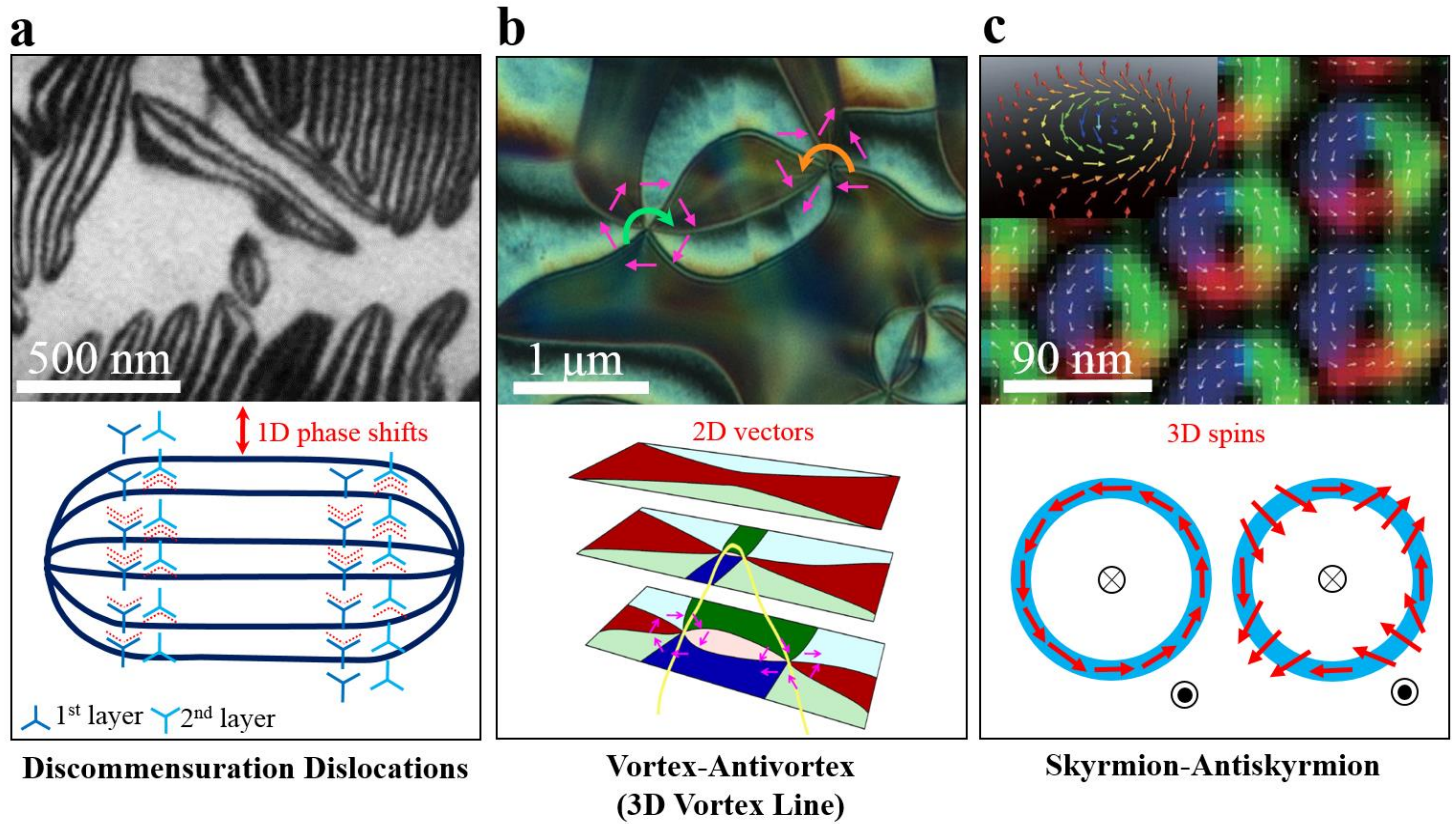

In 2D charge, spin or lattice textures the relevant vectors can be 1D-Ising-type (panel **a**), 2D-*xy*-type (panel **b**), or 3D-Heisenberg-type (panel **c**). The relevant vectors can be phase-shift vectors across domain boundaries or order-parameter vectors/spins in the domains. Panel **a** shows the dark-field transmission electron microscopy (TEM) image of discommensuration dislocations (DDs) in the charge-density-wave state of 2H-TaSe$_2$[59] (the 2H denotes a trigonal prismatic coordination). The six-fold DDs are well studied[59,165,166]: across a discommensuration boundary, a phase shift of ±2/3π (red dotted lines) is observed in the odd layers, but not in the even layers, or vice versa. Even though the total phase-shift change around a DD core is $2\pi$, the phase-shift vectors at the domain boundaries are always one dimensional (that is, scalar), thus DDs cannot be classified as vortices. DDs are not topological defects in the sense that with a 180º flip around an axis in the plane, a DD becomes its counterpart. 1D-shift DDs have been observed in a number of compounds, such as ferroelectric $Ba_2NaNb_5O_{15}$[147,162], ferroelectric $Rb_2ZnCl_4$[163,164] and γ-brass[159-161].

2D vectorial shifts (and 2D vectorial order parameters) are relevant in $Z_m \times Z_n$ domains with topological $Z_l$ vortices. $h$-$RMnO_3$ presents an elegant example (panel **b**): 2D vectorial shifts defined by periodic lattice modulations rotate six times around a vortex core, and a vortex is always connected to antivortices (and vice versa). Vortex and antivortex pairs on a surface are, in fact, a 2D cut of 3D vortex-lines spanning the whole 3D system (panel **b**, yellow line). More examples of defects with 2D relevant vectors can be found in the main text.

3D (pseudo-)vectorial shifts (and 3D (pseudo-)vectorial order parameters) can be found in skyrmions, in which swirling spins form a 3D hedgehog covering the whole spin sphere (panel **c**). Skyrmions are found in noncentrosymmetric magnets[5,167] such as (Fe,Co)Si and $Cu_2OSeO_3$ (panel **c** shows a Lorentz TEM image of $Fe_{0.5}Co_{0.5}Si$)[167]; antiskyrmions also exist[168].

Panel **a** is adapted from REF. 59, panel **c** from REF. 167.

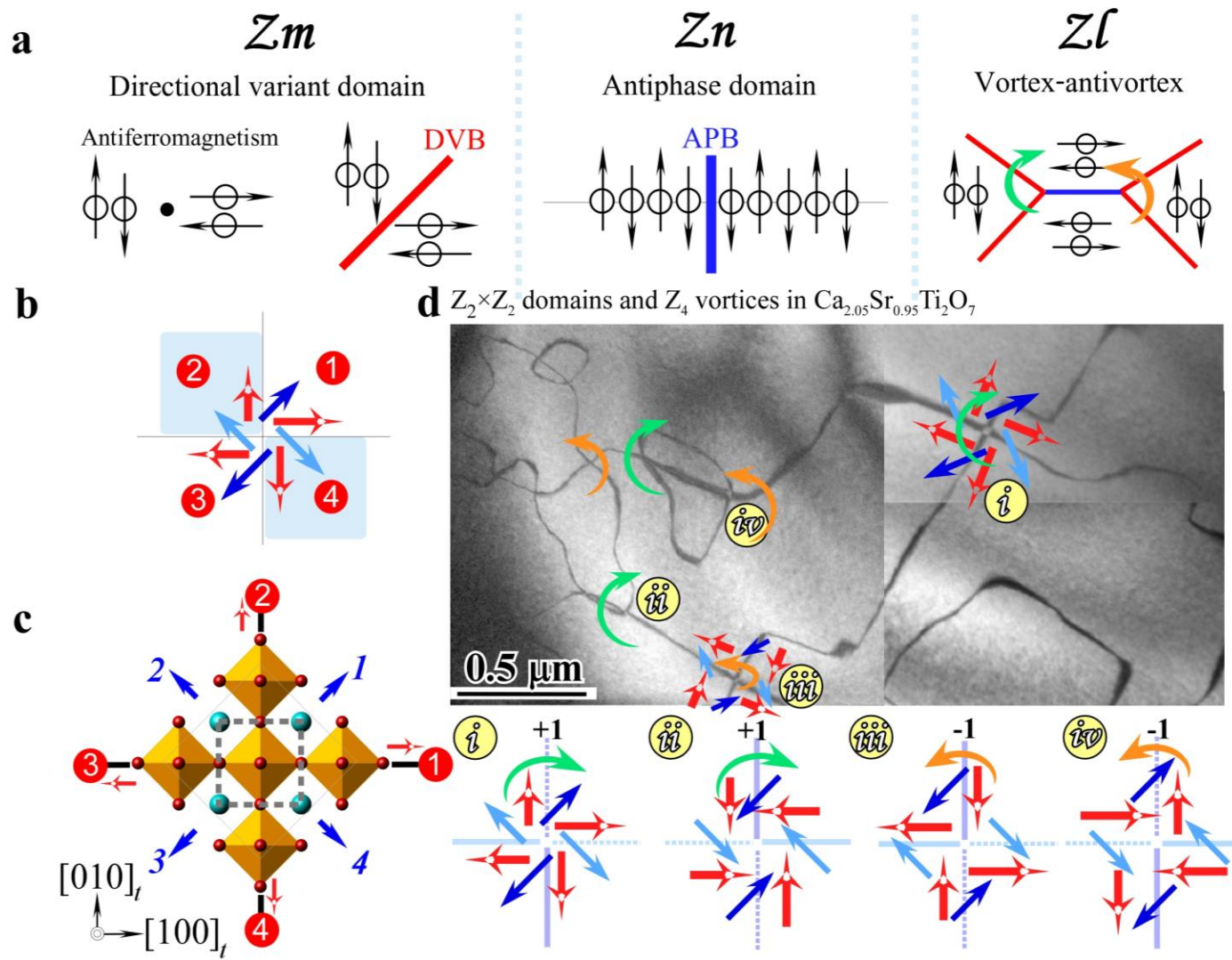

**Fig. 1 $Z_m \times Z_n$ domains and $Z_l$ vortices. a|** Schematic illustrations of antiferromagnetic $Z_2 \times Z_2$ domains and of a $Z_3$ vortex-antivortex pair. Red and blue lines depict directional variant boundaries (DVBs) and antiphase boundaries

(APBs). The green and orange arrows indicate the vorticities obtained by counting the spin directions along the clockwise direction. **b|** A $Z_4$-vortex domain configuration in the bi-layered perovskite $Ca_2Sr_1Ti_2O_7$; the red arrows show apical oxygen tilts, blue and light-blue arrows represent two types of domain boundaries originating from two different tilting axes. **c|** In-plane projected view of $BO_6$ octahedra in the un-distorted Ruddlesden-Popper-series bi-layered perovskite $A_3B_2O_7$. The dashed box outlines the primitive tetragonal cell, and the red and cyan spheres represent O and B ions. Red and blue arrows indicate the eight possible directions of apical oxygen displacements associated with $BO_6$ octahedral tilts. Domains with the $<100>_t$-tilt axes are denoted by the circled numbers, those of the $<110>_t$-tilt axes by the plain numbers. **d|** Dark-field transmission electron microscope images of bilayer $Ca_{2.05}Sr_{0.95}Ti_2O_7$ show $Z_2 \times Z_2$ domains and $Z_4$ vortices. The bottom schematics illustrate four types of topological defects seen in the top image, with oxygen octahedral distortions at domains and domain boundaries within one bilayer. The red arrows denote the domains, the blue and light-blue lines the domain boundaries, and the solid and broken lines are related to each other through a translational $\pi$ shift. The defects can be identified as type i and type ii vortices and type iii and type iv antivortices. Panel **d** is adapted from REF. 50.

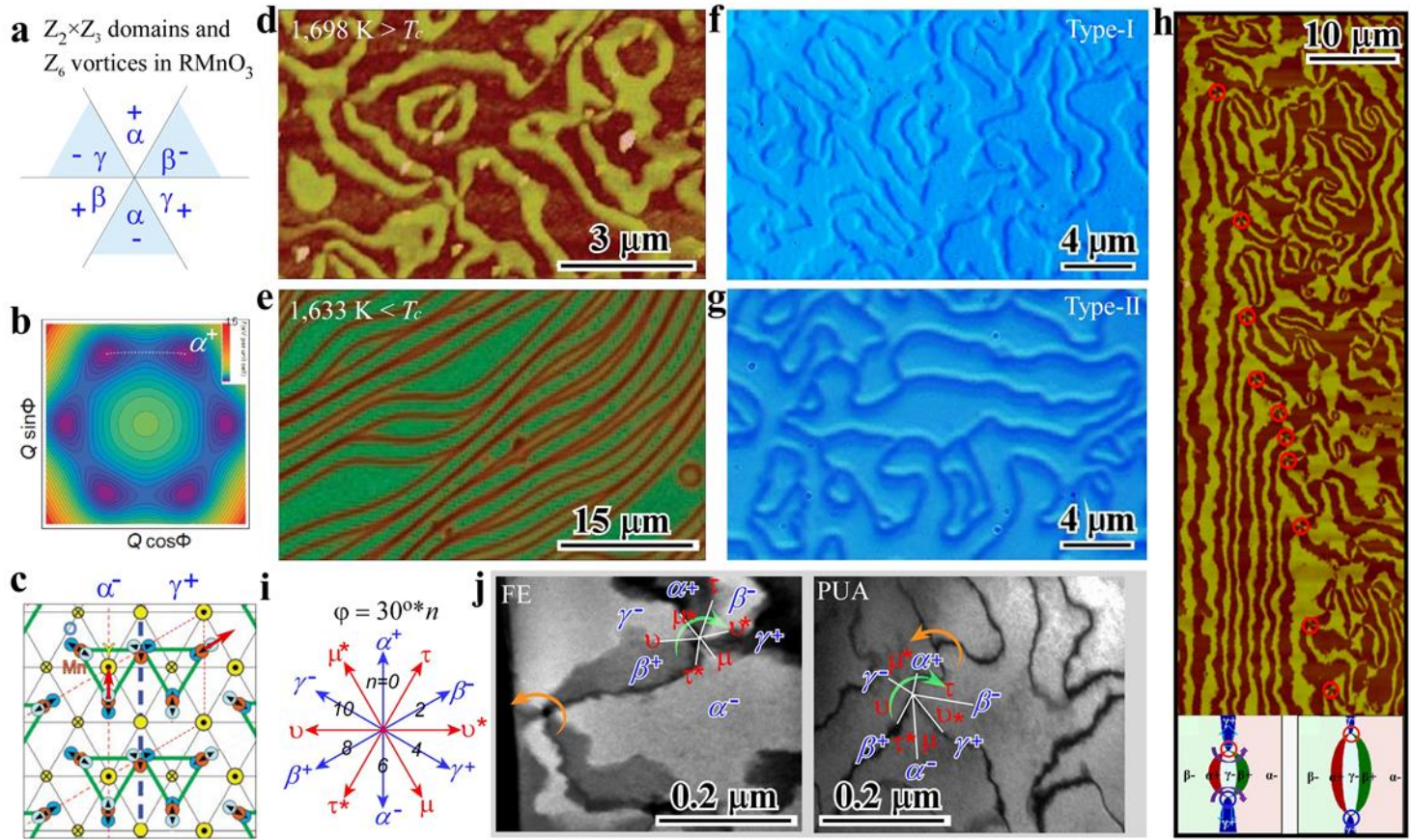

**Fig. 2 $Z_2 \times Z_3$ and $Z_6$ vortices in $h$-RMnO$_3$ a|** A $Z_6$ vortex configuration; directional variants $Z_2$ are associated with + and − ferroelectric polarization, and α, β and γ indicate the three types of antiphase domains. **b|** A contour plot of the free energy of six phases as a function of the trimerization amplitude $Q$ and phase $\varphi$. **c|** A schematic representation of the 60° apical oxygen distortions (red arrows) across the α$^-$–γ$^+$ domain boundaries. The triangles corresponds to the Mn trimers, the grey arrows depict the directions of atomic distortions. **d|** Atomic force microscopy and **e|** optical image of LuMnO$_3$ (critical temperature $T_c$=1,672 K), with the crystal growth or

annealing temperature $T_i$ = 1,698 K for panel d and 1,633 for panel e. Vortices are found for $T_i>T_c$, whereas stripe patterns are observed for $T_i<T_c$. **f,g|** Optical images of YMnO$_3$. Type I domains (f) exhibit almost equal distributions of + and – domains, whereas type II domains (g) display one favoured polarization as a consequence of chemically-driven self-poling. **h|** Atomic force microscopy image of ErMnO$_3$ showing a vortex-to-stripe transformation obtained by applying shear strain with a strain gradient. The bottom panel illustrates the process of pulling apart one vortex-antivortex pair (red-blue circles) using a Magus-type force induced by shear strain. **i|** Twelve possible angles of MnO$_5$ tilts and the corresponding phases. Blue and red arrows correspond to ferroelectric (FE) and partially undistorted antipolar (PUA) states, respectively. **j|** Dark-field transmission electron microscopy images of $h$-InMnO$_3$ (left) and $h$-InMn$_{0.95}$Ga$_{0.05}$O$_3$ (right), showing FE and PUA vortices, respectively. The white lines are the phases at domain boundaries with $n$ odd for a FE vortex, whereas they are the phases at domains for a PUA vortex. Thus, FE and PUA vortices are 'dual' in the exchange of ferroelectricity and antipolarity at domains and domain boundaries. Panel **b** is adapted from REF. 42, panel **c** from REF. 14, panels **d** and **e** from REF. 71, panel **f** and **g** from REF. 54, panel **h** from REF. 107, panel **j** from REFS 52 and 83.

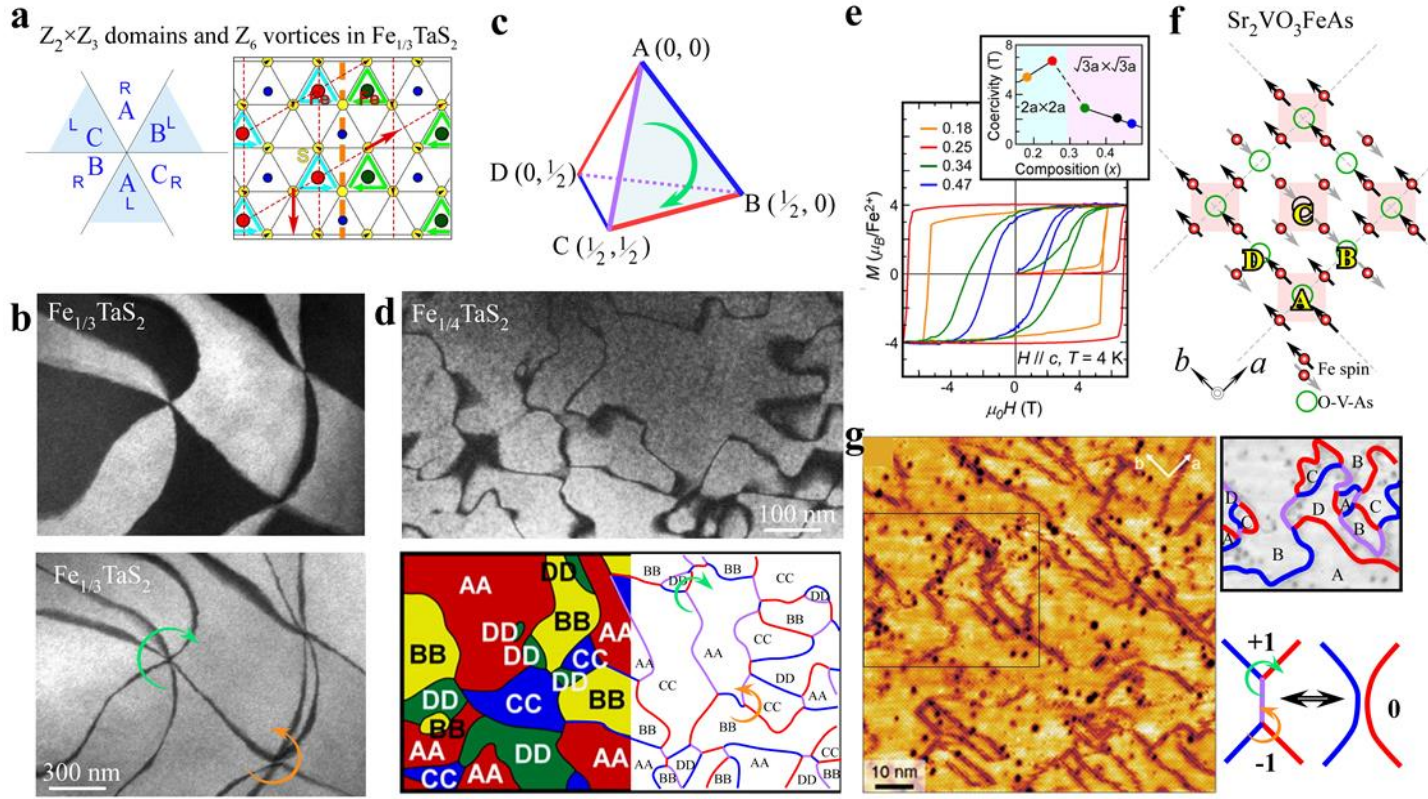

**Fig. 3 $Z_2 \times Z_3$ and $Z_6$ vortices in 2H-Fe$_{1/3}$TaS$_2$ and $Z_1 \times Z_4$ and $Z_3$ vortices. a|** A $Z_6$ vortex configuration and one domain boundary in chiral 2H-Fe$_{1/3}$TaS$_2$. A, B and C represent the three antiphases and R and L denote right and left chirality, which is associated with the rotation of the TaS$_6$ prism (cyan and green triangles). The red arrows indicate the direction of sulfur displacements. **b|** Dark-field transmission electron microscopy (TEM) images of 2H-Fe$_{1/3}$TaS$_2$ measured under different tilting conditions, showing $Z_2$ chiral domains, imaged in contrasting colours (top), and antiphase boundaries (APBs), which appear as dark lines (bottom). **c|** A tetrahedral domain-boundary energy diagram for $Z_1 \times Z_4$ and $Z_3$ vortices. Blue, red and purple edges represent three types of antiphase boundaries,

corresponding to (π, 0), (0, π) and (π, π) phase shifts. **d|** A dark-field TEM image of 2H-Fe$_{1/4}$TaS$_2$ (top) and the corresponding mapping of domains (bottom left) and domain boundaries (bottom right). **e|** Magnetic hysteresis curves of Fe$_x$TaS$_2$ measured at 4 K. The inset shows the magnetic coercivity as a function of composition. **f|** A schematic illustration of the iron spin structure in a plaquette antiferromagnetic order, accompanying four antiphases (A, B, C, D) and three types of domain boundaries (blue, red, purple in panel c). **g|** A spin-polarized scanning tunnelling microscopy image of Sr$_2$VO$_3$FeAs and the domains/domain boundaries mapping. Blue, red and purple domain boundaries always merge at one point. The right panel illustrates the creation/annihilation of a $Z_3$ domain boundary vortex–antivortex pair. Panels **a–c** and **e** are adapted from REF. 56, panel **g** from REF. 61.

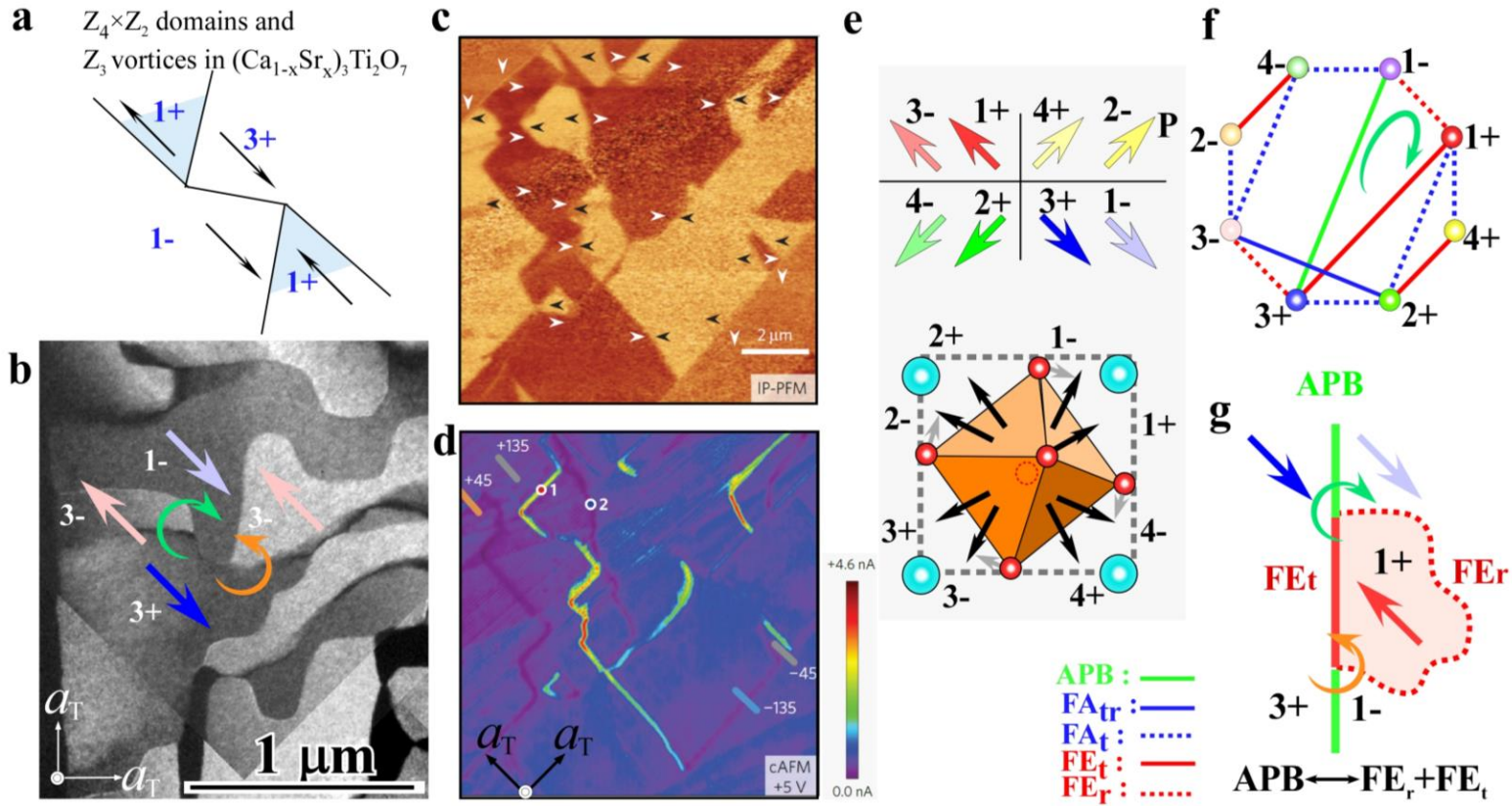

**Fig. 4 $Z_4 \times Z_2$ and $Z_3$ vortices in bilayered perovskite RP327. a|** A $Z_3$ vortex configuration in (Ca,Sr)$_3$Ti$_2$O$_7$. **b|** A mosaic of dark-field TEM (DF-TEM) images of Ca$_{2.55}$Sr$_{0.45}$Ti$_2$O$_7$, showing dark/bright contrast for 180°–type ferroelectric domains within one ferroelastic domain. The arrows indicate polarization directions; domain boundaries of $Z_2$ antiphases (for example, 1- and 3+ domains) are barely visible. **c|** An in-plane piezoresponse force microscopy (IP-PFM) image of the (001) surface of a Ca$_{2.46}$Sr$_{0.54}$Ti$_2$O$_{7-\delta}$ crystal, showing abundant charged ferroelectric boundaries. The arrows indicate the four directions of planar polarization. **d|** The corresponding conductive atomic force microscopy (c-AFM) image shows high conductivity at head-to-head domain boundaries.

**e|** Eight phases and polarization directions are identified based on their characteristic octahedral tilts toward the different quadrants, numbered 1 to 4, and rotations either in the clockwise (+) or anticlockwise (-) direction. **f|** The hyper-tetrahedron energy diagram with eight vertices, representing eight phases (the domains). The eight phases are connected by edges, corresponding to five types of domain boundaries; red and blue edges indicate four types of directional variant boundaries, the green edge one antiphase boundary. **g|** A schematic illustration of a 180º ferroelectric polarization switching through the splitting or coalescence of an antiphase boundary (APB, green) and two ferroelectric walls (FE$_t$, red solid, and FE$_r$, red dashed — FE$_t$ and FE$_r$ boundaries indicate ($X_3^-$)-type tilting and ($X_2^+$)-type rotation directional variant boundaries, respectively), accompanying a $Z_3$ vortex–antivortex creation/annihilation. Panel **d** is adapted from REF. 43, panel **c** from REF. 44.

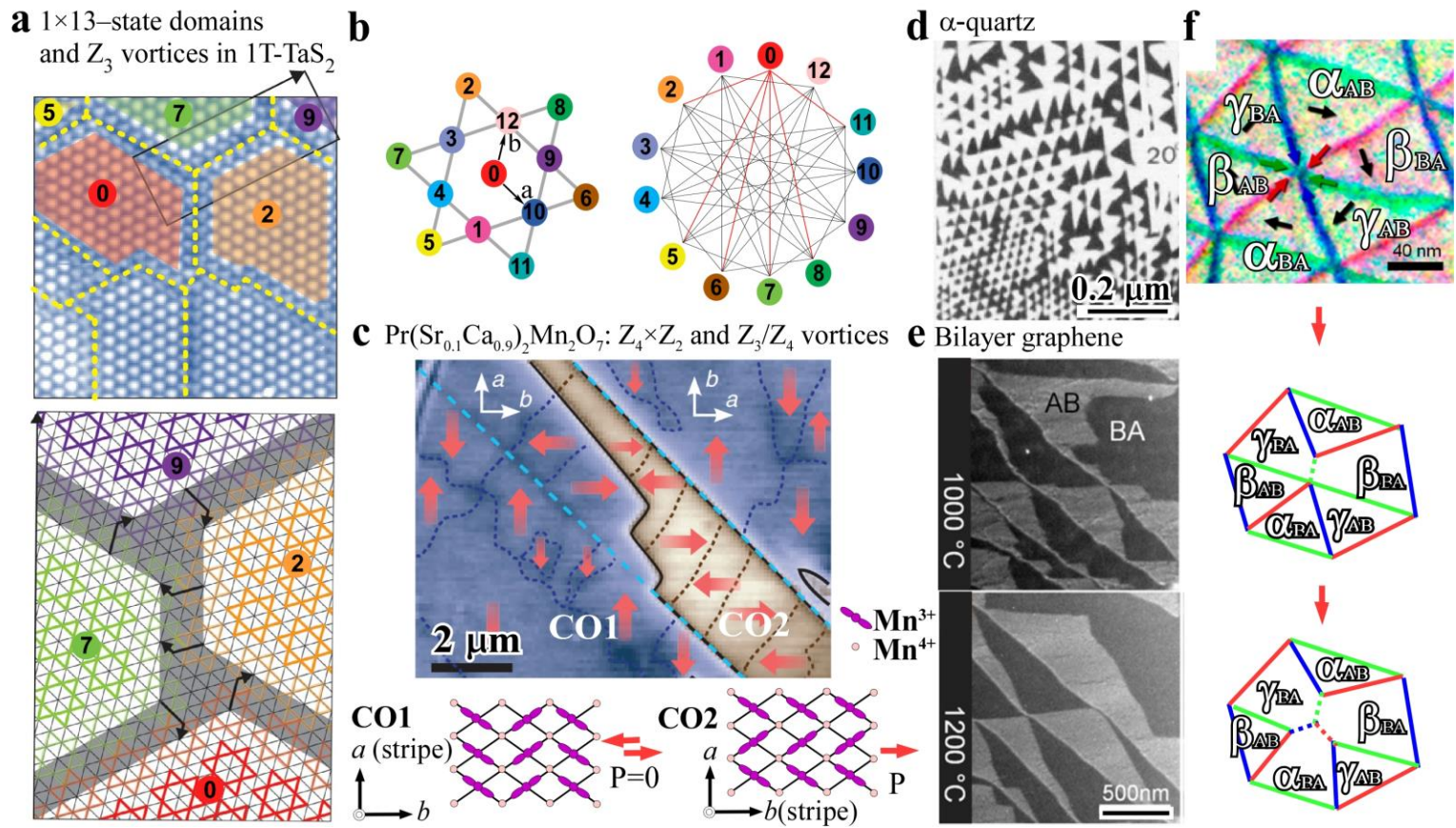

**Fig. 5 Other examples of domain configurations. a|** $1\times13$–state domains and $Z_3$ vortices of 1T-TaS$_2$; the scanning tunnelling microscopy image shows the voltage-pulse-induced charge-density wave domains. All protrusions are identical and constitute David-star-type units of the charge-density wave. The dashed lines denote antiphase boundaries. The bottom panel shows the corresponding domain configurations for the area marked by the rectangular box, exhibiting phase shifts across the antiphase boundaries. **b|** David-star of Ta atoms in the unit cell

with 13 translational origins (left) and the corresponding domain-boundary energy diagram (right), a 'regular graph of 13 vertices with degree 6' with 13 domains and 6 connected edges, which correspond to one distinct ($-a+b$) type of domain boundary (as indicated by the solid edges). **c|** $Z_4 \times Z_2$ and $Z_3$ and $Z_4$ vortices in (anti)ferroelectric states of Pr(Sr$_{0.1}$Ca$_{0.9}$)$_2$Mn$_2$O$_7$. The scanning microwave impedance microscopy image shows charge-order (CO) domain boundaries (dark blue) in the presence of twin boundaries (light blue). The antiferroelectric (CO1) and ferroelectric (CO2) state coexist at 304 K. The white arrows indicate the polarization directions in the 1$^{st}$-bilayer. **d|** Six-fold vertices in a dark-field transmission electron (TEM) microscopy image of Dauphine twins in α-quartz. Upon heating (the sample is cooler at the top and warmer at the bottom), a fragmentation of domains occurs. **e|** Six-fold vertices in a dark-field TEM image of bilayer graphene. Two different stacking sequences (AB and BA) result in two inversion directional variants that give rise to the contrast. The vertices can split upon heating to 1200 °C. **f|** A schematic representation of the possible fragmentation process of a six-fold vertex into four $Z_3$ vortices. Solid lines represent existing antiphase boundaries, dashed lines newly created antiphase boundaries. Panel **b left** is adapted from REF. 58, panel **c** from REF. 51, panel **d** from REF. 157, panels **e, f** from REF. 60.

**Table 1| Techniques for the visualization of various domains in real space.**

| Technique | Resolution* | Specimens | Morphology or physical features | Examples of investigated materials | Refs |
|---|---|---|---|---|---|
| **Surface treatment** | | | | | |
| Chemical etching | cm–μm | not limited | Topography; ferroelectric domains | $h$-R(Mn,Fe)O$_3$ | 54, 71, 96 |
| **Optical methods** | | | | | |
| POM | cm–μm | not limited | Ferroelastic (twins) and chiral domains | $o$-(Ca,Sr)$_3$Ti$_2$O$_7$, BaTiO$_3$ | 3, 27, 54 |

| | | | | $o$-Pr(Sr$_{0.1}$Ca$_{0.9}$)$_2$Mn$_2$O$_7$ | |
|---|---|---|---|---|---|
| SHG | cm–100 nm | not limited | Domains breaking space-inversion or time-inversion symmetry | $h$-R(Mn,Fe)O$_3$, $h$-In(Ga,Mn)O$_3$ | 39, 101, 106, 111 |
| **Scanning probe microscopy** | | | | | |
| PFM | Tens of µm–100 nm | insulators, semiconductors | Electromechanical response; polar domains | $h$-R(Mn,Fe)O$_3$, $h$-In(Ga,Mn)O$_3$, $o$-(Ca,Sr)$_3$Ti$_2$O$_7$, BiFeO$_3$ | 6, 32, 44, 53, 55, 71, 96, 100, 107 |
| c-AFM | Tens of µm–100 nm | semiconductors, conductors | Conductance or tunneling current; charged domain boundaries | BiFeO$_3$, $h$-R(Mn,Fe)O$_3$, $o$-(Ca,Sr)$_3$Ti$_2$O$_7$ | 6, 8, 14, 32, 33, 44, 45 |
| MFM | Tens of µm–20 nm | magnets with net magnetic moments | Tip-specimen magnetic interactions; (weak) ferro(ferri)magnetic domains | LuFe$_2$O$_4$, 2H-Fe$_x$TaS$_2$, $h$-R(Mn,Fe)O$_3$ | 21, 23, 34 |
| MeFM | Tens of µm–20 nm | magnetoelectric insulators | Electric-field-induced magnetoelectricity; magnetoelectric domains | $h$-R(Mn,Fe)O$_3$ | 35 |
| STM | Hundreds of nm–Å | conductors | Tunneling current | 1T-TaS$_2$ | 57, 58 |
| SPSTM | Hundreds of nm–Å | magnetic conductors | Spin-dependent tunneling current; ordered individual spins | Sr$_2$VO$_3$FeAs | 61 |
| sMIM | A few µm–50 nm | semiconductors, conductors | AC electrical conductance in the GHz range | Nd$_2$Ir$_2$O$_7$, $h$-R(Mn,Fe)O$_3$, $o$-Pr(Sr$_{0.1}$Ca$_{0.9}$)$_2$Mn$_2$O$_7$ | 2, 51, 103 |
| **Electron microscopy** | | | | | |
| SEM | Hundreds of µm–10 nm | not limited | Pyroelectric effect and related charging process | $h$-R(Mn,Fe)O$_3$ | 99 |
| DF-TEM | A few µm–15 nm | not limited | Breaking of the Friedel's law; non-centrosymmetric domains and domain boundaries, including antiphase boundaries of ferroelectrics | $h$-R(Mn,Fe)O$_3$, $t'$-Ca$_2$SrTi$_2$O$_7$, $o$-(Ca,Sr)$_3$Ti$_2$O$_7$, 2H-TaSe$_2$, $h$-In(Ga,Mn)O$_3$, 2H-Fe$_x$TaS$_2$, Rb$_2$ZnCl$_4$, Ba$_2$NaNb$_5$O$_{15}$, γ-Brass, graphene, α-quartz | 14, 18, 19, 31, 43, 50, 52, 55, 56, 59, 60, 83, 95, 96, 97, 146, 158, 159, 162, 164 |
| Cs-corrected HRTEM | 50 nm–0.5 Å | not limited | Local structural distortions of domains and domain boundaries, thickness | CaTiO$_3$, Pb(Zr$_{0.2}$Ti$_{0.8}$)O$_3$, $h$-R(Mn,Fe)O$_3$ | 7, 64, 98 |
| Cs-corrected STEM | 10 nm–0.5 Å | not limited | Local structural distortions of domains and domain boundaries, thickness | $h$-R(Mn,Fe)O$_3$, $t'$-Ca$_2$SrTi$_2$O$_7$ $h$-In(Ga,Mn)O$_3$, Pb(Zr$_{0.2}$Ti$_{0.8}$)O$_3$ | 29, 30, 31, 36, 63, 65, 83 |
| Lorentz TEM | A few µm– | magnets with net | Magnetic domains, including skyrmion textures | Fe$_{1-x}$Co$_x$Si, Cu$_2$OSeO$_3$ | 5, 167 |

| | 15 nm | magnetic moments | | | |
|---|---|---|---|---|---|
| **X-ray Imaging** | | | | | |
| X-PEEM | A few µm–hundreds of nm | not limited | Valence state or photo-induced charging; charged domain boundaries | $h$-R(Mn,Fe)O$_3$ | 102 |

*The lateral resolution has been estimated from the references given in the text.

POM: polarized optical microscopy; SHG: second-harmonic generation microscopy; PFM: piezoresponse force microscopy; c-AFM: conductive-atomic force microscopy; MFM: magnetic force microscopy; MeFM: magnetoelectric force microscopy; STM: scanning tunneling microscopy; SPSTM: spin-polarized scanning tunneling microscopy; sMIM; scanning microwave impedance microscopy; SEM: scanning secondary-electron microscopy; DF-TEM: dark-field transmission electron microscopy; HRTEM: high-resolution transmission electron microscopy; STEM: scanning transmission electron microscopy; Lorentz TEM: Lorentz transmission electron microscopy; X-PEEM: X-ray photoemission electron microscopy.

**Table 2| $Z_m \times Z_n$ domains with $Z_l$ vortices.**

| $Z_m \times Z_n$ & $Z_l$ | Examples | Properties of the material | Symmetry operation connecting different $Z_m$ directional variants | $Z_n$ Translational variants (origin of superstructure) | P-state model | Refs |
|---|---|---|---|---|---|---|
| $Z_2 \times Z_2$ & $Z_4$ | $t'$-Ca$_2$SrTi$_2$O$_7$ | Dielectric | $C_2$ rotation | Octahedral tilts ($\sqrt{2}a_p \times \sqrt{2}a_p$) | 4-state Clock | 50 |
| $Z_2 \times Z_3$ & $Z_6$ | h-R(Mn,Fe)O$_3$ | Ferroelectric | Inversion | Mn/Fe trimerization ($\sqrt{3}a_p \times \sqrt{3}a_p$) | 6-state Clock | 14, 31-39, 42, 53-55, 69-71, 94-103, 107 |
| | 2H-Fe$_{1/3}$TaS$_2$ | Chiral | Screw axis | Fe ionic order ($\sqrt{3}a_p \times \sqrt{3}a_p$) | 6-state Clock | 56 |
| | h-In(Ga,Mn)O$_3$ | Antipolar | $C_2$ rotation | Ga/Mn trimerization ($\sqrt{3}a_p \times \sqrt{3}a_p$) | 6-state Clock | 52, 83, 111 |
| $Z_4 \times Z_2$ & $Z_3$ | o-(Ca,Sr)$_3$Ti$_2$O$_7$ | Ferroelectric | $C_4$ rotation | Octahedral tilts & rotations ($\sqrt{2}a_p \times \sqrt{2}a_p$) | 8-state Potts | 43, 44 |
| $Z_1 \times Z_4$ & $Z_3$ | 2H-Fe$_{1/4}$TaS$_2$ | Ferromagnetic | Identity | Fe ionic order ($2a_p \times 2a_p$) | 4-state Potts | 23, 56 |
| | Sr$_2$VO$_3$FeAs | Antiferromagnetic (superconducting) | Identity | Plaquette AFM order ($2a_p \times 2a_p$) | 4-state Potts | 61 |
| $Z_4 \times Z_2$ & $Z_3$ /$Z_4$ | o-Pr(Sr$_{0.1}$Ca$_{0.9}$)$_2$Mn$_2$O$_7$ (CO1) | Antiferroelectric | $C_4$ rotation | Chare/orbital stripes ($1a_p \times 2a_p$) | 8-state Potts | 51 |
| | o-Pr(Sr$_{0.1}$Ca$_{0.9}$)$_2$Mn$_2$O$_7$ (CO2) | Polar | | Charge/orbital stripes ($2a_p \times 1a_p$) | | |

| | | | | | | | |
|---|---|---|---|---|---|---|---|
| *1x13* & $Z_3$ | 1T-TaS$_2$ | Charge-density wave | Identity | David star-type CDW ($\sqrt{13}a_p \times \sqrt{13}a_p$) | 13-state extended Potts | 57, 58 | |

1T and 2H indicate two polytypes of transition metal dichalcogenides; *h*: hexagonal; *t'*: tetragonal; *o*: orthorhombic; $a_p$: primitive lattice, AFM: antiferromagnetic, CDW: charge-density wave.